\documentclass[prl,letterpaper,twocolumn,amssymb,nofootinbib]{revtex4}
\usepackage{graphicx}
\newcommand{\oml}{\Omega_{\Lambda}}
\newcommand{\etal}{{\it et al.}}
\newcommand{\AP}{Alcock-Paczy\'nski }
\begin{document}

\title{Testing the Cosmic Coincidence Problem and the Nature of Dark Energy}

\author{Neal Dalal, Kevork Abazajian, Elizabeth Jenkins and Aneesh V. Manohar}

\affiliation{Department of Physics, University of California at San Diego,
La Jolla, CA 92093-0319}

\begin{abstract}
\bigskip
Dark energy models which alter the relative scaling behavior of dark
energy and matter could provide a natural solution to the cosmic
coincidence problem - why the densities of dark energy and dark matter
are comparable today.  A generalized class of dark energy models is
introduced which allows non-canonical scaling of the ratio of dark
matter and dark energy with the Robertson-Walker scale factor a(t).
Upcoming observations, such as a high redshift supernova survey,
application of the Alcock-Paczynski test to quasar pairs, and cluster
evolution, will strongly constrain the relative scaling of dark matter
and dark energy as well as the equation of state of the dark energy.
Thus, whether there actually is a coincidence problem, and the extent
of cosmic coincidence in the universe's recent past can be answered
observationally in the near future.  Determining whether today is a
special time in the history of the universe will be a SNAP.
\end{abstract}

\maketitle

Recent observations of supernov\ae\ \cite{p99}, CMB anisotropies \cite{cmb} and
large scale structure \cite{lss,bahcall} point to the presence of a flat
universe with a dark energy component.\footnote{Since the distinction between
dark matter and dark energy is that the latter has (negative) pressure, perhaps
a more appropriate term would be the Dark Force.} The exact nature of the dark
energy is not clear, but it must contribute a significant fraction of the
closure density today, and have sufficiently negative pressure to
cause an acceleration of the  
Hubble expansion rate at recent times. A cosmological constant of magnitude
$\oml\equiv\rho_\Lambda/\rho_{\rm crit}\approx 0.7$ 
provides an excellent fit to the experimental data, and so
the flat $\Lambda$CDM cosmology has become the leading cosmological model.

However, $\Lambda$CDM is beset by several serious theoretical difficulties,
which may be characterized as fine-tuning problems. Experimental data require
that the vacuum energy density today be of order $\sim (10^{-3} {\rm eV})^4$,
whereas the natural scale for the vacuum energy density, theoretically, is of
order $m_{Pl}^4$, $m_W^4$, or at best $\Lambda_{\rm QCD}^4$. Thus some unknown
physics must fine-tune $\Lambda$ by 40--120 orders of magnitude below its
natural value. This problem is called the cosmological constant
problem~\cite{lambdarefs}. Another related but distinct difficulty with
$\Lambda$CDM is the so-called ``why now?'' or coincidence problem.  Briefly
put, if $\Lambda$ is tuned to give $\oml\sim\Omega_M$ today, then for
essentially all of the previous history of the universe, the cosmological
constant was negligible in the dynamics of the Hubble expansion, and for the
indefinite future, the universe will undergo a de Sitter-type expansion in
which $\oml$ is near unity and all other components are negligible.  The
present epoch would then be a very special time in the history of the universe,
the only period when $\Omega_M\sim\oml$.

The cosmological constant and coincidence problems have led numerous authors to
consider alternatives to $\Lambda$CDM which preserve its stunning successes
(Type Ia SNe, CMB anisotropies, large-scale structure) but avoid the above
difficulties.  The modification which involves the smallest departure from
conventional thinking is to postulate that the acceleration of the universe is
due not to a cosmological constant, but instead to an evolving component with
sufficiently negative equation of state; typically one uses a scalar field
\cite{tracker,quintessence}.  The expansion of the universe in such models can
mimic that of a universe with a cosmological constant. These models typically
fail to address the coincidence problem since matter density and vacuum energy
density evolve differently, and are only comparable for some period of time,
which is chosen to coincide with the current epoch. To see this, note that if
matter and dark energy are coupled only gravitationally, then they are
conserved separately, so the matter density scales as $\rho_M\propto a^{-3}$
while the dark energy density scales as $\rho_X\propto a^{-3(1+w_X)}$, taking
the equation of state parameter $w_X=p_X/\rho_X$ to be constant.  Thus, the
ratio $\rho_M/\rho_X\propto a^{3w_X}$.  Fairly negative equations of state are
required to explain the supernova data \cite{ptw}, $w_X\sim -1$.  Matter
rapidly was therefore dominant over dark energy at even moderate
redshift, leading 
to the coincidence problem.  This argument holds true even if $w_X$ evolves
with redshift. As long as $w_X$ is sufficiently negative for recent
times, dark energy is dynamically important only today. Equivalently, the total
equation of state $w_{\rm tot}=w_X \Omega_X\approx 0$ in the recent past,
even if $w_X$ evolves with redshift away from $-1$.  To circumvent
the coincidence problem requires a more radical departure from conventional
cosmology, such as assuming that there is some mechanism relating the effective
cosmological constant to the matter density at all times.  Several proposed
theories possess this property.  Such theories involve modifications of
gravitation, or nonminimal coupling between dark energy and
matter~\cite{sss,carroll}.

Little is known about the relation between vacuum and matter energy densities
over the history of the universe. It is not experimentally known whether there
is a coincidence problem, since we have no experimental information on the
variation of $\Omega_\Lambda$ and $\Omega_M$ with time for the recent history
of the universe. In the simplest $\Lambda$CDM scenario, in which the vacuum
energy is indeed due to a cosmological constant, the densities scale as 
\begin{equation} 
\rho_X \propto  \rho_M a^3.
\end{equation}
In a theory with no coincidence problem, one expects
\begin{equation} 
\rho_X \propto  \rho_M.
\end{equation}
Other theoretical scenarios proposed in the literature entail vastly different
evolutions of matter and energy density.  It seems premature at this stage to
do a detailed fit of data to a particular model (e.g. a scalar field with a
particular form for the potential).   Rather, we suggest the alternative
approach of using the experimental data to constrain the nature of dark energy
with minimal underlying theoretical assumptions. A useful starting point is to
assume a phenomenological form for the ratio of the dark energy and matter
densities (valid from some redshift $z_{\rm max}$ till today),
\begin{equation}\label{xidef} 
\rho_X \propto  \rho_M a^\xi, \hbox{ i.e. } \Omega_X \propto \Omega_M a^\xi,
\end{equation}
where the scaling parameter $\xi$ is a new observable. The special cases
$\xi=3$ and $\xi=0$ correspond to $\Lambda$CDM and the self-similar solutions
\cite{sss}, respectively. The value of $\xi$ quantifies the severity of the
coincidence problem, and can be constrained by several means, three of which we
describe below.

We will assume a flat universe, $\Omega_M+\Omega_X=1$, throughout. This is not
essential, but simplifies the expressions (and is an observational fact
\cite{cmb}).  Energy conservation requires
\begin{equation}
\frac{d\rho_{\rm tot}}{da}+\frac{3}{a}(1+w_X \Omega_X)\rho_{\rm tot}=0,
\label{energycons}
\end{equation}
where $\rho_{\rm tot}=\rho_M+\rho_X$ is the total density, which gives
\begin{equation}
\frac{\rho_{\rm tot}}{\rho_0}=\exp\left[\int_a^1 {da\over a} 3  
(1+w_X \Omega_X)\right].
\label{rhototeqn}
\end{equation}
Taking $w_X$ constant gives
\begin{equation}
\rho_{\rm tot}=\rho_0a^{-3}[1-\Omega_{X,0}(1-a^\xi)]^{-3w_X/\xi}
\label{modeleqn}
\end{equation}
where $\Omega_{X,0}$ is the value of $\Omega_X$ today. For $\xi=0$, the
solution is $\rho_{\rm tot}=\rho_0 a^{-\beta}$, with $\beta=3(1+w_X \Omega_X)$.

This family of solutions, with three parameters ($\Omega_{X,0},w_X,\xi$),
allows us to parameterize a wide range of possible cosmologies in a simple
fashion.  $\Omega_{X,0}$ specifies the current density in dark energy, $w_X$
specifies its equation of state, and $\xi$ specifies how strongly
$\Omega_X/\Omega_M$ varies with redshift. With this parameterization in hand,
we can explore how well future observations can constrain the time evolution of
dark energy density, and thereby limit models of dark energy.  As we shall
show, many of the classical tests proposed to determine the existence and
magnitude of the cosmological constant also are well-suited for testing the
coincidence problem in dark energy models.  Limiting ($\Omega_{X,0},w,\xi$) is
a more efficient procedure than trying to individually constrain the
multitudinous theoretical models proposed in the literature. In addition, the
parameters have a simple  interpretation, and two important special cases,
$\Lambda$CDM and self-similar models, are included as $\xi=3$ and $\xi=0$
models, respectively.

\begin{figure}
\includegraphics[width=0.45\textwidth]{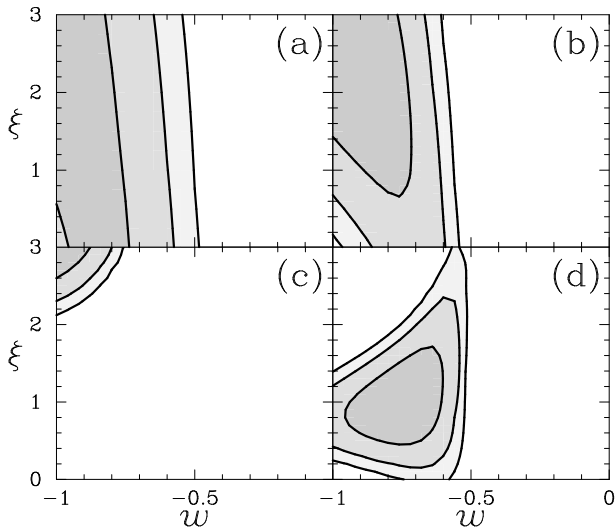}
\caption{Parameter extraction with future supernova surveys.  Panel (a) shows
likelihood contours in the ($w,\xi$) plane for the currently known 42 high
redshift SNe \cite{p99}.  Panel (b) illustrates the capabilities of a ground
based supernova survey which observes 350 SNe up to redshift $z=1.2$, taking an
input $\Lambda$CDM cosmology.  Panel (c) shows likelihood contours for 6000 SNe
between redshifts $0.3\leq z<1.7$, corresponding to 3 years of SNAP data, again
using an input $\Lambda$CDM cosmology.  Panel (d) shows the same calculation as
(c), but instead using an input cosmology of $\Omega_{X,0}=0.7$, $w=-0.75$,
$\xi=1$.  Contours correspond to confidence levels of 68\%, 95\%, and 99\%.  In
all cases we marginalize the likelihood over $\Omega_{X,0}$ using a Gaussian
prior centered on $\Omega_{X,0}=0.7$ with width of 0.1.\label{ml}}
\end{figure}

The first constraint we consider is the redshift-luminosity relation of
high-redshift Type Ia supernov\ae.  The luminosity distance $d_L(z)$ to
supernov\ae\ is given by
\begin{eqnarray}
d_L(z)&=&c(1+z)\int_{(1+z)^{-1}}^1\frac{da}{a^2 H}\\\nonumber
&=&\frac{c}{H_0}(1+z)\int_0^z dz (\rho_{\rm tot}/\rho_0)^{1/2}
\end{eqnarray}
where $\rho_{\rm tot}$ is given by Eq.~(\ref{rhototeqn}). The
redshift-magnitude relation for the 42 moderate redshift SNe presented in
Ref.~\cite{p99} is unable to distinguish between standard $\Lambda$CDM and a
self-similar model with constant $\oml$, as shown in Fig.~\ref{ml}(a). Future
observations \cite{snap} will result in more stringent constraints, as
discussed below.

To determine how well high redshift SNe may be used to distinguish a given set
of parameters ${\vec p}$ from the true parameters ${\vec p}_0$, we construct
$\chi^2$
\begin{eqnarray}\label{chi2}
\chi^2({\vec p},{\vec p}_0) &=& \frac{N}{\sigma^2}\bigg(\int_0^{z_m}
[f({\vec p},{\vec p}_0,z)]^2 W(z)\frac{dz}{z_m} \\\nonumber
&& - \left[\int_0^{z_m}
f({\vec p},{\vec p}_0,z)W(z)\frac{dz}{z_m}\right]^2\bigg),
\end{eqnarray}
\begin{equation}
f({\vec p},{\vec p}_0,z)=5\log_{10}\left[\frac{d_L({\vec
p},z)}{d_L({\vec p}_0,z)}\right].
\end{equation}
Here, $N$ is the total number of SNe, $\sigma\approx 0.2$ is the error in the
observed supernova magnitudes, $z_m$ is the maximum redshift out to which SNe
are followed, and $W(z)$ is a weighting function that describes the redshift
distribution of the detected SNe.  Note that the second term in Eq.
(\ref{chi2}) arises because the supernova absolute magnitude  is a free
parameter in the fit \cite{p99}. For simplicity, we take the redshift
distribution to be uniform, $W(z)=1$.

Fig.~\ref{ml}(b--d) displays parameter likelihoods computed using Eq.
(\ref{chi2}).  We first consider a ground-based survey, capable of detecting
350 SNe up to redshift $z=1.2$.  Such a survey would improve the current limits
on parameters $w$ and $\xi$ only marginally,  disfavoring self-similar models
relative to the input $\Lambda$CDM model at the $\sim1\sigma$ level, as shown
in panel (b).  We note that this estimate is probably optimistic, since the
errors in the SNe magnitudes are likely to be larger than what we have assumed
at the high redshift end, where the greatest leverage on $\xi$ is possible. 
The SNAP satellite, on the other hand, offers the definitive answer to the
question of whether there is a coincidence problem.  SNAP should observe
roughly 6000 SNe up to redshifts $z<1.7$ over its 3-year lifetime, which would
allow an unequivocal determination of the presence or absence of a coincidence
problem, as shown in panels (c) and (d).  We note that most of SNAP's ability
to measure $\xi$ derives from the highest redshift SNe, so there is incentive
to follow these events to even higher redshifts ($z\sim 2$) than is currently
planned.

\begin{figure}
\includegraphics[width=0.45\textwidth]{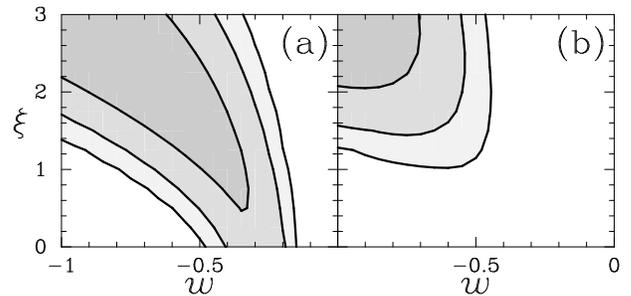}
\caption{Parameter extraction with the \AP test.  Panel (a) shows likelihood
contours (68\%, 95\%, and 99\% respectively)  derived by placing 25 quasar
pairs randomly in redshift with a Gaussian distribution centered on $z=2$ with
width $\Delta z=1$.  We use Eq. (6) of Ref.~\cite{hui} to evaluate the ability
of the AP test to distinguish various cosmologies from $\Lambda$CDM.  In panel
(b) we show the joint likelihood  (same confidence levels as panel (a)) of the
AP test along with the currently known 42 high redshift SNe. The combined
constraints severely limit possible models, and lead to excellent determination
of $\xi$.  In both cases we marginalize over $\Omega_{X,0}$ assuming the same
prior as in Fig. \ref{ml}.
\label{ap}}
\end{figure}

SNAP is scheduled to launch late this decade, however other methods may be
employed in the interim. Besides high redshift SNe, another means of testing
the coincidence problem is the \AP (AP) test~\cite{ap,hui,jordi}.  This test
distinguishes cosmologies by measuring the product of $H(z) d_A(z)$, where
$H(z) = {\dot a}/a$ is the Hubble parameter, and $d_A=d_L/(1+z)^2$ is the
angular diameter distance.  Following Ref. \cite{hui,jordi}, we consider an
implementation of this method using the Lyman $\alpha$ forest.  If redshift
distortions due to peculiar velocities are negligible, then we can analytically
estimate the ability of this test to distinguish between cosmologies
\cite{hui}.  We place 25 quasar pairs randomly in redshift, with a Gaussian
distribution centered on $z=2$, assume separations of $1^\prime$, and evaluate
the ability of the \AP test to distinguish various models from the input
$\Lambda$CDM model.  We plot in Fig. \ref{ap} the likelihood contours in
$(w,\xi)$ space, again marginalized over $\Omega_{\Lambda,0}$ using a Gaussian
prior centered on $\Omega_{\Lambda,0}=0.7$ with width 0.1.  Although the \AP
test can rule out a significant class of models, a large region of parameter
space is nearly degenerate with 25 quasar pairs.  However, the combination of
the AP test with other constraints can dramatically shrink the allowed region. 
In panel (b) of Fig. \ref{ap} we plot the joint likelihood obtained by
combining constraints from the currently known 42 high redshift SNe \cite{p99}
with the 25 quasar pairs plotted in panel (a).  Although the current SNe data
do not constrain $\xi$, when they are combined with the AP constraints, strong
limits can be placed on the evolution of $\Omega_X/\Omega_M$.  Since the \AP
test can already be performed today, and the Sloan Digital Sky Survey will
provide large numbers of quasar pairs in the imminent future (e.g.
\cite{sdss}), this method is a promising technique to probe the cosmic
coincidence question.

A final example we consider is the evolution of cluster abundances at
high redshifts.  As is well known, cluster abundances place strong
constraints on cosmology; for example, a single $M\approx
10^{15}M_\odot$ cluster at $z>0.8$ rules out $\Omega_M=1$ with high
significance \cite{bahcall} assuming Gaussian initial fluctuations.
Similarly, galaxies observed at high-redshift appear to have formed
earlier than predicted by leading models \cite{peebles}.  Our models
with $\xi<3$ predict even less evolution than $\Lambda$CDM, because
the dark energy remains dynamically important further into the past.
Therefore, the presence or absence of clusters at high redshifts
$(z>1)$ can place strong constraints on the coincidence problem.

\begin{figure}
\includegraphics[width=0.45\textwidth]{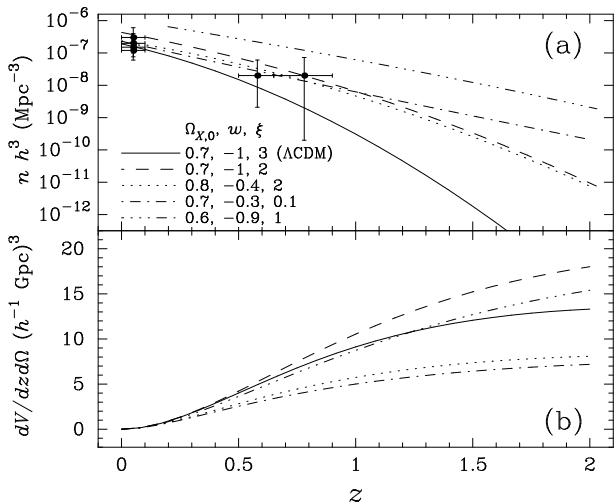}
\caption{Other methods.  In panel (a) we plot the predicted number density of
massive clusters for different parameter choices. The self-similar models
predict far less evolution between $z\sim 2$ and the present time than
$\Lambda$CDM.  Data points with errorbars are taken from Ref. \cite{fan}.  In
panel (b), we plot comoving volume elements for the same models plotted in the
first panel. Note that nearly degenerate models in panel (a) are well separated
in panel (b).\label{other}}
\end{figure}

We use the Press-Schechter approximation to estimate the rate of cluster
evolution with redshift; see Ref. \cite{fan} for details.   To compare with
observations, we plot the number density not as a function of the virial mass,
but as a function of the mass enclosed within $1.5 h^{-1}$ Mpc, by computing
the virial overdensity $\Delta$  using the spherical collapse model \cite{eke}
and rescaling the mass using the profile $M(<r)\propto r^{0.64}$ \cite{fan}. 
For simplicity, we assume that observed clusters collapse and virialize at the
redshift at which they are observed. Examples of cluster evolution are plotted
in the first panel of  Fig.~\ref{other}.  As with the other tests discussed
above, cluster evolution can clearly discriminate between self-similar models
and tracker models.  However, it appears that  there will be degenerate regions
of parameter space.

Besides the examples described in this work, other probes of the coincidence
problem will be possible.  At moderate redshifts, number counts of galaxies
\cite{newman} (see Fig.~\ref{other}) or gravitational lensing statistics
\cite{kochanek} should strongly constrain the evolution of dark energy, if
systematic effects such as evolution can be accurately modeled.  At higher
redshifts, constraints on $\Omega_X$ may be derived from Big Bang
nucleosynthesis at $z\approx 10^6$ or CMB anisotropies at $z\approx 10^3$
\cite{bean}.  Given that the dark energy is still so poorly understood, it is
not clear whether these high redshift limits may be extrapolated to recent
epochs, when the coincidence problem arises.  One possible constraint
from CMB anisotropies on the recent evolution of dark energy is the
normalization of the power spectrum.  Normalization using $\sigma_8$ and COBE
are consistent for standard, low density models, however
our models with $\xi<3$ and $w\sim -1$ could ruin this consistency by
changing the angular power spectrum on large scales ($l\sim 10$) due
to the integrated Sachs-Wolfe effect.

We have shown that several independent methods may be used to test the
coincidence problem, as well as test whether the simplest dark energy scenarios
are feasible or whether radical departures from conventional thinking are
required.  All of these measurements are either currently feasible (and in
progress), or will soon be forthcoming.  Of the methods we discuss, the most
definitive conclusions will be possible using the SNAP satellite, which
hopefully will fly later this decade.  SNAP's ability to
constrain the evolution of $\Omega_X$ will be enhanced by following SNe
at very high redshift, $z\gtrsim 2$.  In the near term, strong constraints may
be placed on dark energy using cluster evolution or the \AP test.  We note that
if these latter methods give preliminary hints of a departure from standard
($\xi+3w=0$) cosmologies, then it will be imperative to build experiments like
SNAP to further study dark energy.

We thank Eric Gawiser, Nao Suzuki, Wallace Tucker and Art Wolfe for
helpful discussions. This work was supported in part by the U.S.\
Dept.\ of Energy, under grant DOE-FG03-97ER40546.  KA was partially
supported by NASA under GSRP, and ND acknowledges support from the
ARCS Foundation.

\bigskip\hrule
\end{document}